\documentclass[preprint,prb,superscriptaddress,amssymb,amsmath,floatfix,citeautoscript]{revtex4-1}
\usepackage[pdftex]{graphicx}
\usepackage{natbib}

\begin{document}

\title{Heat dissipation in the quasiballistic regime studied using Boltzmann equation in the spatial frequency domain}
\author{Chengyun Hua}
\affiliation{Oak Ridge National Laboratory, Oak Ridge, TN 37831, USA}
\author{Austin J. Minnich\footnote{\text{
To whom correspondence should be addressed. E-mail: aminnich@caltech.edu}}}
\affiliation{%
California Institute of Technology, Pasadena, California 91125,USA
}

\date{\today}

\begin{abstract}

Quasiballistic heat conduction, in which some phonons propagate ballistically over a thermal gradient, has recently become of intense interest. Most works report that the thermal resistance associated with nanoscale heat sources is far larger than predicted by Fourier's law; however, recent experiments show that in certain cases the difference is negligible despite the heaters being far smaller than phonon mean free paths. In this work, we examine how thermal resistance depends on the heater geometry using analytical solutions of the Boltzmann equation. We show that the spatial frequencies of the heater pattern play the key role in setting the thermal resistance rather than any single geometric parameter, and that for many geometries the thermal resistance in the quasiballistic regime is no different than the Fourier prediction. We also demonstrate that the spectral distribution of the heat source also plays a major role in the resulting transport, unlike in the diffusion regime. Our work provides an intuitive link between the heater geometry, spectral heating distribution, and the effective thermal resistance in the quasiballistic regime, a finding that could impact strategies for thermal management in electronics and other applications.

\end{abstract}

\maketitle

\section{Introduction}
Quasiballistic heat conduction occurs if a temperature gradient exists over length scales comparable to phonon mean free paths (MFPs)\cite{Chen,Majumdar1993,hua_transport_2014,Cahill:2014}. In this regime, local thermal equilibrium does not exist and Fourier's law is no longer valid. Heat dissipation under these conditions is of intense interest because the thermal resistance from a heat source may be far greater than predicted by Fourier's law, impacting microelectronics and other applications. 

Many recent experiments have investigated quasiballistic thermal transport using optical pump-probe methods to heat a metal transducer film or lithographically patterned nanoscale heaters. Siemens \emph{et. al.} were the first to perform these experiments using an optical pump and soft x-ray probe.\cite{Siemens2010} Numerous other experiments have been reported including using transient grating\cite{Johnson2013}, thermoreflectance methods\cite{Minnich2011PRL,ding_radial_2014, Cahill2007, Wilson2014}, and other lithographically patterned metallic heaters.\cite{zeng_measuring_2015,hu_spectral_2015,hoogeboom-pot_new_2015}

In general, these experiments report that the thermal transport from heaters of characteristic length smaller than phonon MFPs is increasingly impeded compared to the Fourier's law prediction as the heater size decreases. However, recent experiments have shown that situation may be more complicated than the conventional viewpoint. Hoogeboom-Pot \emph{et al} probed the thermal transport in sapphire and silicon using arrays of nickel nanowires as heat sources and showed that when the separation between nanoscale line heaters was small compared to the dominant phonon MFPs, the measured thermal boundary resistance recovered to the diffusive limit.\cite{hoogeboom-pot_new_2015} Our recent work used aluminum nanoline arrays as heating sources in TDTR experiments and showed that the measured thermal conductivity of sapphire reached a constant even as the linewidth decreased.\cite{Chen2017} These experiments suggest that the actual dependence on the heater dimensions and geometry cannot simply be attributed to a single characteristic dimension of the heater. 

Many theoretical frameworks have been developed to explain the observed quasiballistic phenomena. Vermeersch \emph{et. al.} reported a truncated levy formalism to analyze the quasiballistic motion of thermal energy in semiconductor alloys.\cite{Vermeersch2015a, Vermeersch2015b} Two-channel models, in which low and high frequency phonons are described by the Boltzmann transport equation (BTE) and heat equation, have been applied to analyze the quasiballistic thermal transport in transient grating\cite{Maznev2011} and thermoreflectance experiments\cite{Wilson2013,Dames2015}. A McKelvey-Shockley flux method was adapted to describe phonon transport.\cite{Maassen2015} Analyses within the framework of the BTE have been recently reported in a number of studies.\cite{Regner2014, Ramu2014, Collins2013APL, hua_transport_2014} Most of the works simplified the BTE either by assuming a gray model or by asymptotic expansion of spatial part of the phonon distribution. Zeng and Chen numerically solved a gray BTE to analyze the quasiballistic heat conduction trends for periodic nanoscale line heaters. They found opposite trends in the effective thermal conductivity depending on the measures of temperature profiles used to extract the thermal conductivities.\cite{zeng_disparate_2014} Despite all of these works, a unified explanation of the reported data for thermoreflectance experiments with different heater geometries remains lacking.

In this work, we analyze heat dissipation in the quasiballistic regime using an analytical solution to the mode-dependent BTE in the spatial frequency domain. Our study shows that the thermal resistance is not a function of any one geometrical parameter but rather depends on the spatial frequencies of the heater pattern and their relative weights. The use of spatial frequencies differs from most prior works that considered geometrical parameters like linewidths as the key parameters. Our work provides an intuitive link between the heater geometry, spectral distribution of the heat source, and the effective thermal resistance in the quasiballistic regime, a finding that could impact strategies for thermal management in electronics and other applications.

\section{Theory}
The transport of heat by thermal phonons is described by the Boltzmann transport equation (BTE) under the relaxation time approximation given by\cite{Majumdar1993}
\begin{equation} \label{eq:BTE_energy}
\frac{\partial g_{\omega}}{\partial t}+\mathbf{v}_g \cdot \nabla g_{\omega} = -\frac{g_{\omega}-g_0(T)}{\tau_{\omega}}+\frac{Q_{\omega}}{4\pi},
\end{equation}
where $g_{\omega} = \hbar\omega D(\omega) (f_{\omega}(\mathbf{r},t,\theta,\phi)-f_0(T_0))$ is the desired deviational distribution function, $g_0(T)$ is the equilibrium deviational distribution function defined below, $Q_{\omega}(\mathbf{r},t)$ is the spectral volumetric heat generation, $\mathbf{v}_g(\omega)$ is the phonon group velocity, and $\tau_{\omega}$ is the phonon relaxation time. Here, $\mathbf{r}$ is the spatial vector, $t$ is the time, $\omega$ is the phonon frequency, and $T(\mathbf{r},t)$ is the local temperature. In the Cartesian coordinate system and assuming an isotropic crystal, the advection term in Eq.~(\ref{eq:BTE_energy}) is expanded as
\begin{equation}\label{eq:advection}
\mathbf{v}_g \cdot \nabla g_{\omega} = v_g \mu \frac{\partial g_{\omega}}{\partial z}+ v_g \sqrt{1-\mu^2}\text{cos}\phi \frac{\partial g_{\omega}}{\partial x}+ v_g \sqrt{1-\mu^2}\text{sin}\phi \frac{\partial g_{\omega}}{\partial y},
\end{equation}
where $\mu = cos(\theta)$ is the directional cosine of the polar angle $\theta$ and $\phi$ is the azimuthal angle. Here, we note that while many crystals like Silicon contain minor anisotropies in the Brillouin zone, they remain thermally isotropic.

Assuming a small temperature rise, $\Delta T(\mathbf{r},t) = T(\mathbf{r},t) - T_0$, relative to a reference temperature $T_0$, the equilibrium deviational distribution is proportional to $\Delta T(\mathbf{r},t)$, 
\begin{equation}
g_0(T) = \frac{1}{4\pi}\hbar \omega D(\omega) (f_{BE}(T) - f_{BE}(T_0)) \approx \frac{1}{4\pi}C_{\omega}\Delta T(\mathbf{r},t).
\label{eq:BEDist_Linearized}
\end{equation}
Here, $\hbar$ is the reduced Planck constant, $D(\omega)$ is the phonon density of states, $f_{BE}(T)$ is the Bose-Einstein distribution, and $C_{\omega} = \hbar\omega D(\omega)\frac{\partial f_{BE}}{\partial T}$ is the mode specific heat. The volumetric heat capacity is then given by $C = \int_0^{\omega_m}C_{\omega}d\omega$ and the thermal conductivity $k = \int_0^{\omega_m}k_{\omega}d\omega$, where $k_{\omega} = \frac{1}{3} C_{\omega}v_{\omega} \Lambda_{\omega}$ and $\Lambda_{\omega} = \tau_{\omega}v_{\omega}$ is the phonon MFP. 

Both $g_{\omega}$ and $\Delta T$ are unknown. Therefore, to close the problem, energy conservation is used to relate $g_{\omega}$ to $\Delta T$, given by 
\begin{equation}
\int\int_0^{\omega_m} \left[\frac{g_{\omega}(\mathbf{r},t)}{\tau_{\omega}}-\frac{1}{4\pi}\frac{C_{\omega}}{\tau_{\omega}}\Delta T(\mathbf{r},t) \right]d\omega d\Omega = 0,
\label{eq:EnergyConservation}
\end{equation}
where $\Omega$ is the solid angle in spherical coordinates and $\omega_m$ is the cut-off frequency. Note that summation over phonon branches is implied without an explicit summation sign whenever an integration over phonon frequency is performed. 

By assuming an infinite domain in the in-plane directions ($x$ \& $y$) and a semi-infinite domain in cross plane direction ($z$) with specular boundary conditions such that the domain can be extended to infinity by symmetry, in Ref. 18 we derived the Green's function of BTE in Fourier space, given by: 
\begin{equation}\label{eq:Temperature_FourierTransform}
H(\eta,\xi_x,\xi_y,\xi_z) = \frac{\int^{\omega_m}_0\frac{Q_{0\omega}}{\Lambda_{\omega}\xi}\text{tan}^{-1}\left(\frac{\Lambda_{\omega}\xi}{1+i\eta\tau_{\omega}}\right) d\omega}{\int^{\omega_m}_0 \frac{C_{\omega}}{\tau_{\omega}}\left[1-\frac{1}{\Lambda_{\omega}\xi}\text{tan}^{-1}\left(\frac{\Lambda_{\omega}\xi}{1+i\eta\tau_{\omega}}\right)\right]d\omega},
\end{equation}
where $\xi = \sqrt{\xi^2_x+\xi^2_y+\xi^2_z}$ is the spatial Fourier variable and $\eta$ is the temporal Fourier variable. Note that $Q_{0\omega}$ is the volumetric heating spectral profile, a scalar with units of $Q_{\omega}(\mathbf{r},t)$ in Eq.~(\ref{eq:BTE_energy}) that depends on phonon frequency. The temperature response to any input is now simply the product of Eq.~(\ref{eq:Temperature_FourierTransform}) and the input function $\bar{Q}(\eta,\xi_x,\xi_y,\xi_z)$, given by
\begin{equation}\label{eq:TempResponse}
\Delta \widetilde{T}(\eta,\xi_x,\xi_y,\xi_z) = H(\eta,\xi_x,\xi_y,\xi_z)\times \bar{Q}(\eta,\xi_x,\xi_y,\xi_z),
\end{equation}
where the volumetric heat input is then given as  $Q = \int^{\omega_m}_0Q_{0\omega}d\omega \bar{Q}(\eta,\xi_x,\xi_y,\xi_z)$. 

This equation can be used to examine the thermal resistance of a semi-infinite slab to heater patterns of different geometries and characteristic dimensions. Note that we do not explicitly include an interface that is present in thermoreflectance experiments. Despite this simplification, our results are still useful to understand these experiments because the interface only changes in spectral distribution of the heater, not the trends with spatial and temporal frequencies.

In the calculations that follow, we use the dispersion and lifetimes for crystalline silicon calculated by Lucas Lindsay using density functional theory.\cite{Lindsay2013} The details regarding converting the ab-initio calculations to an isotropic dispersion can be found in Ref. 18.\nocite{hua_analytical_2014} 

\begingroup
\squeezetable
\begin{table}
\centering
\begin{tabular}{lc}
\hline
\hline
Heating pattern & Function \\
\hline
Gaussian spot & $\delta(\eta-\eta_0)\frac{d}{1+d^2\xi_z^2}e^{-\frac{D^2\xi^2_r}{8}}$ where $\xi^2_r = \xi^2_x+\xi^2_y$ \\
Line array & $\delta(\eta-\eta_0)\frac{d}{1+d^2\xi_z^2} \sum\limits_{n = -\infty}^{+\infty} \frac{\text{sin}(n\pi \frac{w}{L})}{n}$ (n $\in$ Integers)\\
Dot array & $ \delta(\eta-\eta_0)\frac{d}{1+d^2\xi_z^2}\sum\limits_{n = -\infty}^{+\infty} \sum\limits_{m = -\infty}^{+\infty} \frac{\text{sin}(n\pi \frac{w}{L})\text{sin}(m\pi \frac{w}{L})}{mn}$ (n, m $\in$ Integers)\\
\hline
\hline
\end{tabular}
\caption{Input function $\bar{Q}$ in Fourier space for Gaussian spot heating, nanoline array heating, and nanodot array heating. For the Gaussian heater, $D$ is the $1/e^2$ width of a Gaussian distribution, also called the Gaussian diameter. In the nanoline and nanodot array heating patterns, the ratio of the linewidth or dot width $w$ and the period $L$, $w/L$, is defined as duty cycle. The heating is assumed to be periodic in time with frequency $\eta_0$ and exponentially decaying in the cross-plane direction with a decay depth $d = 10$ nm. }
\label{table:heatingfunction}
\end{table}
\endgroup

\begin{figure*}[t!]
\centering
\includegraphics[scale = 0.3]{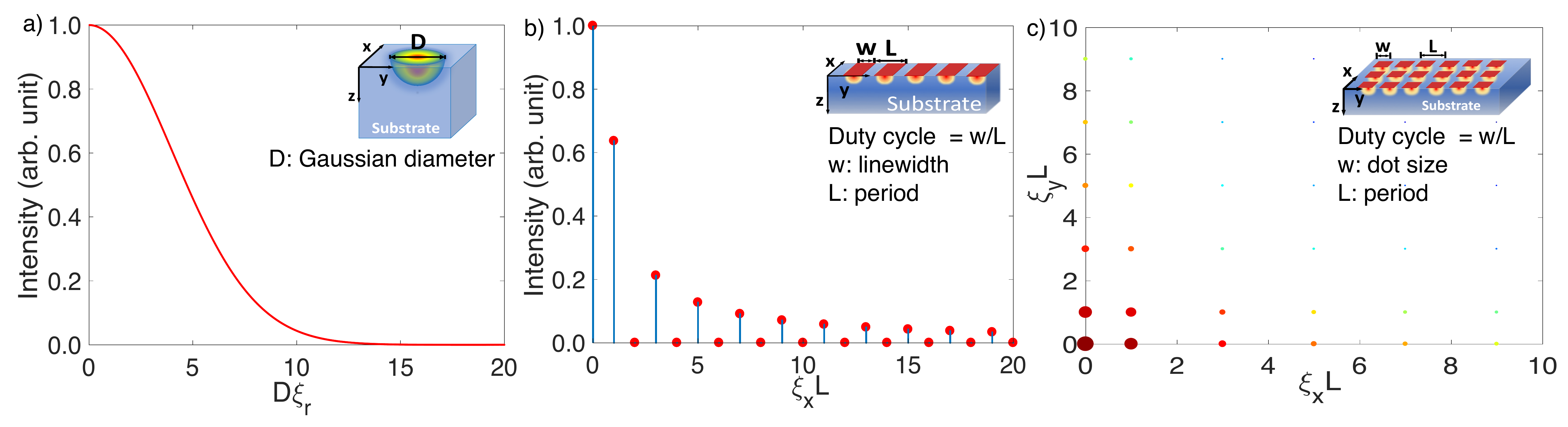}
\caption{Input function $\bar{Q}$ versus in-plane spatial frequency of (a) a Gaussian distribution, (b) nanoline arrays and (c) nanodot arrays. The functions have a maximum at zero in-plane spatial frequency and decrease with increasing spatial frequency. In (c), the size and color of the dots indicate the magnitude of the Fourier component at that spatial frequency.   Insets: real-space schematics of each heater pattern.}
\label{fig:Geometry}
\end{figure*}

\section{Results}

We begin by examining thermal transport for the three heating geometries used in recent experiments: a Gaussian distribution, nanoline arrays, and nanodot arrays. The spectral heating distribution $Q_{0\omega}$ is set to be $A C_{\omega}/\tau_{\omega}$ where $A$ is a constant indicating the magnitude of the heat input, which we denote as a thermalized distribution. The Fourier transform of the three heating input functions is tabulated in Table \ref{table:heatingfunction} and plotted in Fig.~\ref{fig:Geometry}(c). Except for the discrete nature of line and dot arrays, the Fourier components of all three functions have two common features: their maximum occurs at zero spatial frequency, and the non-zero elements decrease monotonically as $\xi_x$ or $\xi_y$ increases. 

\begin{figure*}[!]
\centering
\includegraphics[scale = 0.5]{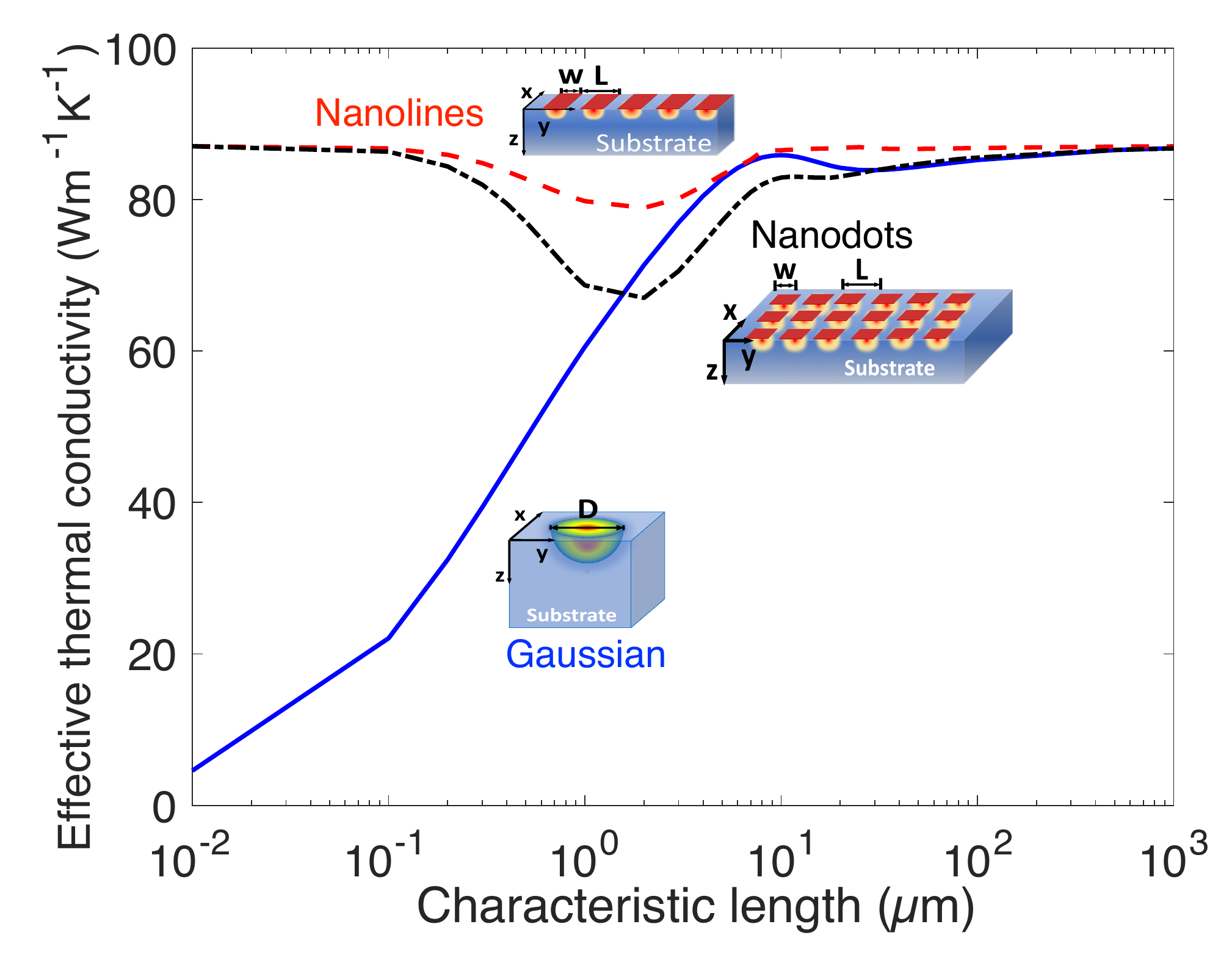}
\caption{ Effective thermal conductivity versus characteristic length for three heater patterns at a fixed temporal frequency $\eta_0 = 1$ MHz. The characteristic length corresponds to diameter $D$ for the Gaussian heater (solid line), period $L$ for the nanoline (dashed line) and nanodot (dotted-dash line) array heaters with a 50\% duty cycle. The effective thermal conductivity decreases as $D$ decreases while it exhibits a non-monotonic trend as $L$ decreases for the nanoline and nanodot arrays. }
\label{fig:Effectivek}
\end{figure*}

Despite the similarities in the input functions, the thermal resistances exhibit dissimilar trends. We obtain a measure of the thermal resistance by calculating the surface temperature rise for the three heating patterns at a fixed temporal frequency $\eta_0 = 1$ MHz using the following equation
\begin{equation}\label{eq:SurfaceResponse}
\Delta T(x,y,\eta_0) = \int\int^{\infty}_{-\infty}H_s(\xi_x,\xi_y,\eta_0)Q(\xi_x,\xi_y)e^{i\xi_x x}e^{i\xi_y y} d\xi_x d\xi_y,
\end{equation}
where $H_s(\xi_x,\xi_y,\eta_0) = \int^{\infty}_{-\infty}H(\eta_0,\xi_x,\xi_y,\xi_z)\frac{d}{1+d^2\xi_z^2}d\xi_z$ is the surface temperature response at the surface, and $\bar{Q}(\xi_x,\xi_y)$ is the input function that only involves the in-plane heating geometry. The surface temperature rise is then fitted to a diffusion model based on Fourier's law where the thermal conductivity is treated as a fitting parameter. In this way, we can extract effective thermal conductivities, which provide an indication of the overall resistance to heat flow. Note that if the in-plane spatial frequencies are zero then the heater is a thin film. 

The results are shown in Fig.~\ref{fig:Effectivek}. There are three major features in this figure. First, when the characteristic length $L_c$ is such that $\Lambda_{\omega}L_c \ll 1$, the effective thermal conductivity of all three heaters approaches a constant value, 89 Wm$^{-1}$K$^{-1}$, which is less than the bulk value, 130 Wm$^{-1}$K$^{-1}$, because of the cross-plane effects which will be discussed shortly. Second, for the Gaussian heater, when the beam diameter decreases and hence in-plane spatial frequency increases, the effective thermal conductivity indeed decreases monotonically, as has been reported in experiment\cite{Minnich2011PRL, Wilson2014}. However, the same calculation applied to a line array heater or a dot array heater shows that the effective thermal conductivity initially decreases as period decreases, and then increases back to the constant value as the period further decreases. Third, the dot array heater has a larger drop in effective thermal conductivity compared to the line array heater. 

In the following sections, we will analyze and explain the reasons behind these observed features. 

\subsection{Cross-plane quasiballistic effects and volumetric heating spectral profile}

\begin{figure*}[t!]
\centering
\includegraphics[scale = 0.24]{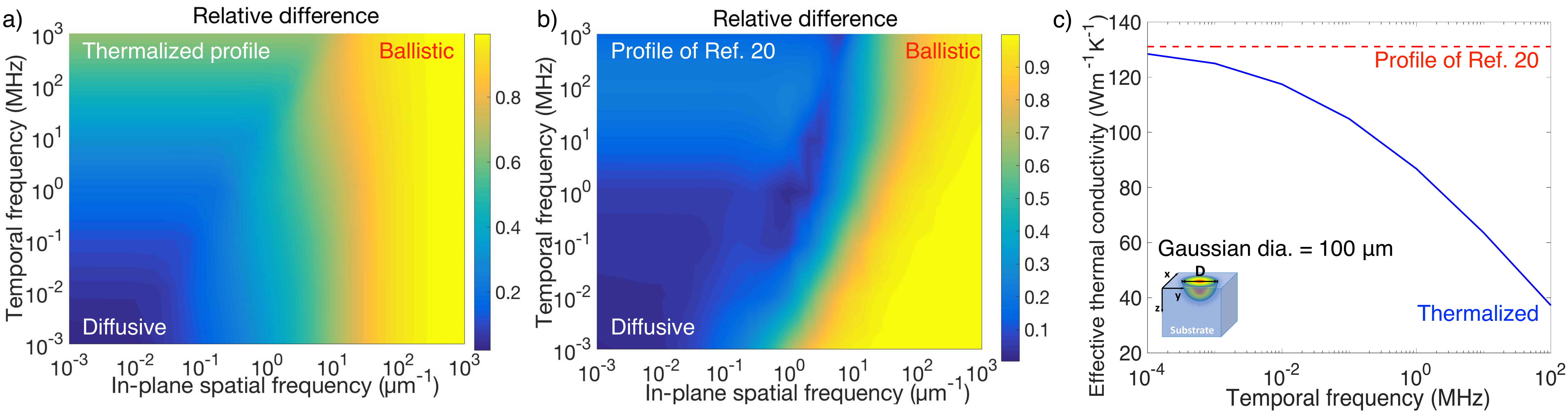}
\caption{Relative difference between BTE and Fourier predicted surface temperatures, $|\frac{H_{sBTE}-H_{sFourier}}{H_{sBTE}}|$ versus temporal frequency $\eta$ and in-plane spatial frequency $\xi_x$ for (a) the thermalized heating spectral profile, $Q_{0\omega} = C_{\omega}/\tau_{\omega}$ and (b) the spectral profile given in Ref. 20. Lighter colors indicate larger deviations between the BTE and Fourier calculations. Deviations occur as spatial and temporal frequencies increase. (c) Effective thermal conductivity versus temporal frequency of a Gaussian heater  with the thermalized volumetric heating profile (solid line) and the profile of Ref. 20 (dashed line). }
\label{fig:ContourH}
\end{figure*}

First, we examine the role of cross-plane effects. Consider the surface response $H_s(\xi_x,\xi_y,\eta)$ with the thermalized spectral profile as shown in Fig.~\ref{fig:ContourH}(a). The contour plot shows the relative difference between BTE and Fourier's law surface responses, $|\frac{H_{sBTE}-H_{sFourier}}{H_{sBTE}}|$ as a function of temporal frequency $\eta$ and in-plane spatial frequency $\xi_x$. Deviations from Fourier's law are observed at $\xi_x$ = 0.1 $\mu$m$^{-1}$ corresponding to a period of 10 $\mu$m. Note that deviations are observed even if $\xi_x$ = 0 for temporal frequencies corresponding to the cross-plane effects.

Consider the origin of the cross-plane effects at $\xi_x = 0$. To observe quasiballistic effects, one of the following conditions must be satisfied: $\eta\tau_{\omega} \sim 1$, or $\xi_z\Lambda_{\omega} \sim 1$. At temporal frequencies less than a few hundred MHz, $\eta\tau \ll 1$ as the relaxation times are mostly less than a few nanoseconds. Therefore, the origin of the cross-plane effects must be related to the cross-plane spatial frequencies $\xi_z$. Recall that an oscillating surface heat flux induces a thermal wave with a characteristic depth given by $\sqrt{k\eta^{-1}C^{-1}}$. At megahertz frequencies, the spatial frequencies corresponding to this characteristic depth satisfy $\xi_z\Lambda_{\omega} \sim 1$, and hence deviations from Fourier's law are observed.  For example, at a temporal frequency 1 MHz, the thermal penetration depth in silicon is around 10 $\mu$m, comparable to phonon MFPs in silicon, which range from a few nanometers to tens of micrometers. 

This cross-plane effect is strongly influenced by the spectral profile of the heating, $Q_{0\omega}$. As shown in Fig.~\ref{fig:ContourH}(b), the deviation due to $\xi_z$ can be eliminated by choosing a different spectral profile, such as that of phonons injected across an interface from Ref. 20.\nocite{Hua2017} Such dependence does not occur in the diffusion regime. The origin of this dependence can be see in the numerator of Eq.~\ref{eq:Temperature_FourierTransform}: the spectral volumetric heating profile $Q_{0\omega}$ is modified by $\frac{1}{\Lambda_{\omega}\xi}\text{tan}^{-1}\left(\frac{\Lambda_{\omega}\xi}{1+i\eta\tau_{\omega}}\right)$. This dependence enables transmission coefficients to be extracted from TDTR measurements as described in Ref. 20. 

As a result, even if $\xi_z\Lambda \sim 1$, a cross-plane effect in the effective thermal conductivity may not be observed. For example, Fig.~\ref{fig:ContourH}(c) shows that in a Gaussian heater with a diameter of 100 $\mu$m, the effective thermal conductivity for the thermalized spectral profile exhibits a decreasing trend versus temporal frequency, while for the profile of Ref. 20 the effective thermal conductivity remains constant. 

Here, the spectral profile of the heat source can significantly alter the effective thermal resistance of the solid to heat flow. Even though the responses between BTE and Fourier's law appear to agree with each other, it does not necessarily mean quasiballistic effects are absent. To avoid confusion, in the rest of the analysis we will use the thermalized profile. 

\subsection{In-plane quasiballistic effects}

We now consider the trends in effective thermal conductivity with in-plane spatial frequency for each heater. Neither the surface response $H_s(\xi_x,\xi_y)$ nor the input function $\bar{Q}(\xi_x,\xi_y)$ alone can explain the thermal conductivity trends in Fig.~\ref{fig:Effectivek}. However, the overall surface response as given in Eq.~\ref{eq:SurfaceResponse} is an integral of the product of $H_s(\xi_x,\xi_y)$ and $\bar{Q}(\xi_x,\xi_y)$ over in-plane spatial frequencies, weighted by the factor $e^{i\xi_x x}e^{i\xi_y y} d\xi_x d\xi_y$.  Therefore, understanding the observed trends requires examining this weighted product of $H_s(\xi_x,\xi_y)$ and $\bar{Q}(\xi_x,\xi_y)$ versus in-plane spatial frequency. 

Without loss of generality, $x$ and $y$ are set to zero for simplicity. To separate the in-plane effects from the cross-plane effects, this weighted product of $H_s$ and $Q$ is divided by $H_s(\xi_x= 0,\xi_y= 0)$, the DC component of the surface temperature response. We define $H_sQd\xi_x d\xi_y/H_s(\xi_x=0,\xi_y= 0)$ as a normalized response $\bar{H}_s$. In this way, the difference between BTE and Fourier's normalized responses $\bar{H}_s$ is solely caused by the in-plane effects, and the DC components of BTE and Fourier's $\bar{H}_s$ are always identical. 

\begin{figure*}[t!]
\centering
\includegraphics[scale = 0.3]{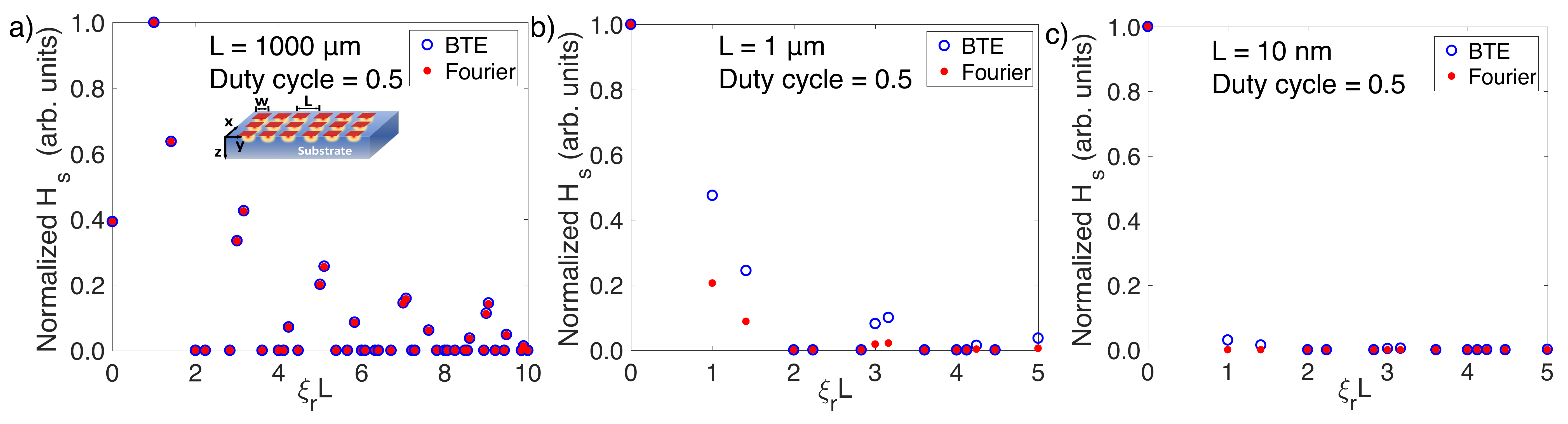}
\caption{ Normalized response $\bar{H}_s(\xi_r,\eta)$ versus non-dimensional in-plane spatial frequency for a dot array heater with a duty cycle at 50 \% and a period $L$ of (a) 1000 $\mu m$, (b) 1 $\mu m$, and (c) 10 nm. The BTE solution (open circles) is compared to the Fourier's law prediction (solid circles). }
\label{fig:WeightedHs_Dot}
\end{figure*}

The normalized response $\bar{H}_s$ versus non-dimensional in-plane spatial frequency for a dot array heater with a duty cycle at 50 \% and various period are shown in Fig.~\ref{fig:WeightedHs_Dot}. As in Fig.~\ref{fig:WeightedHs_Dot}(a), if $\xi_r \Lambda_{\omega} \ll 1$, then the BTE and Fourier weighted responses are identical. As the spatial frequency increases, the BTE solution deviates from the Fourier's solution and thus we would expect the weighted response to deviate as well. This behavior is indeed observed for the dot array heater with a period of 1 $\mu m$ as shown in Fig.~\ref{fig:WeightedHs_Dot}(b). 

As the spatial frequency further increases, the deviation in the BTE response continues to increase as well. However, the magnitude of the responses becomes much smaller than that of the DC component.  Therefore, the DC component dominates the overall thermal response, and as discussed earlier the BTE and Fourier normalized responses $\bar{H}_s$ are identical at zero in-plane spatial frequency. As a result, even when $\Lambda_{\omega}\xi_r \gg 1$, the surface response of the dot array heaters is identical to that of a dot array heater with a large period. The effective thermal conductivity returns to its constant value set by the cross-plane effects, yielding the dip trend shown in Fig.~\ref{fig:Effectivek}. Similar reasoning explains the trend of the effective thermal conductivity in the line array heaters.

\begin{figure*}[t!]
\centering
\includegraphics[scale = 0.42]{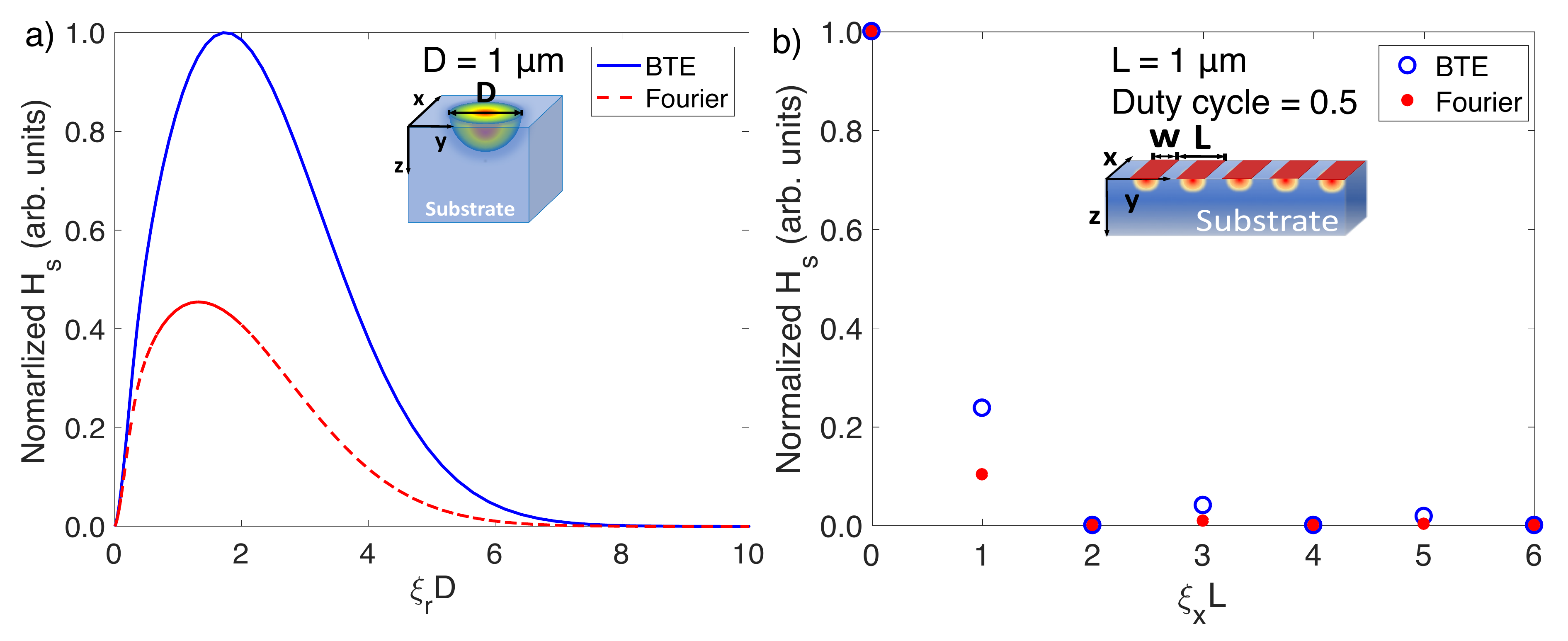}
\caption{Normalized response $\bar{H}_s(\xi_r,\eta)$ versus non-dimensional in-plane spatial frequency for (a) a Gaussian heater with a Gaussian diameter $D$ at 1 $\mu m$ and (b) a line array heater with a period $L$ at 1 $\mu m$ and a duty cycle at 50 \%. The BTE solution (solid line in (a); open circles in (b)) is compared to the Fourier's law prediction (dashed line in (a); solid circles in (b)).}
\label{fig:WeightedHs_GaussianAndLinearray}
\end{figure*}

On the other hand, for the Gaussian heater the effective thermal conductivity keeps decreasing as the Gaussian diameter decreases. The reason can be identified from Fig.~\ref{fig:WeightedHs_GaussianAndLinearray}(a). The amplitude of the normalized response $\bar{H}_s$ in a Gaussian heater is maximum at a non-zero in-plane spatial frequency, and the DC component is always zero. The peak in the response shifts to a bigger in-plane spatial frequency as the Gaussian diameter decreases. As a result, the deviation in the surface response of the Gaussian heaters between BTE and Fourier's law keeps increasing, leading to a decreasing trend of the effective thermal conductivity. 

The key distinguishing feature of these three cases can be attributed to the relative contribution of the DC component. In a one-dimensional problem, such as the line arrays, the integral to obtain the surface response is performed along a single dimension, for instance $\xi_x$. The DC component will always make a major contribution because the amplitude of the normalized response $\bar{H}_s$ scales as $\xi_x^{-1}$. In the continuous two-dimensional problem, the integral in Eq.~\ref{eq:SurfaceResponse} is performed over $\xi_x$ and $\xi_y$. Rewriting $d\xi_xd\xi_y$ in polar coordinates as $\xi_r d\xi_rd\phi$, where $\xi_r = \sqrt{\xi_x^2+\xi_y^2}$ and $\phi$ is the polar angle, we observe that the DC component contributes nothing to the integral when the spatial frequencies are continuous, as occurs for an isolated heater like the Gaussian spot. Hence, the deviations in thermal response between the BTE and Fourier solutions are readily observable, as in Fig.~\ref{fig:Effectivek}. 

For a two-dimensional periodic heat source like the dot arrays, the DC response returns but higher harmonics may still impact the overall response. The major difference between a line array and a dot array heater is that the harmonics in a line array heater are composed of two points ($\pm 2\pi/L$, $\pm 4\pi/L$, \emph{etc}) while the harmonics in a dot array heater are composed of at least four points, \emph{i.e.} the first harmonic is composed of points at ($\pm 2\pi/L$, 0) and (0,$\pm 2\pi/L$). Therefore, the relative contribution from the harmonics in a dot array heater is bigger than that in a line array heater due to their higher degeneracy, leading to a bigger drop in the effective thermal conductivity in dot array heaters as shown in Fig.~\ref{fig:Effectivek}.

\begin{figure*}[t!]
\centering
\includegraphics[scale = 0.42]{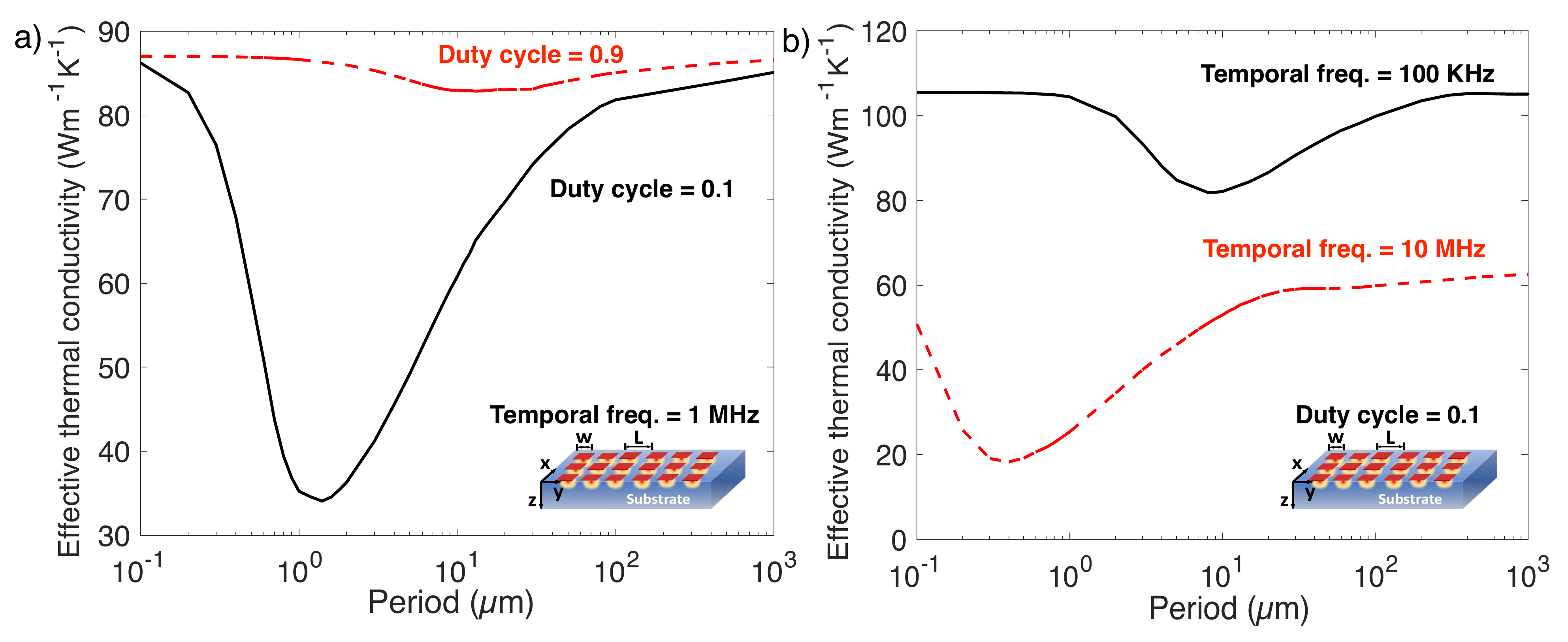}
\caption{(a) Effective thermal conductivity versus period with a duty cycle at 90 \% (dashed line) and 10 \% (solid line). (b)  Effective thermal conductivity versus period with a duty cycle of 10 \% at temporal frequencies 100 KHz (solid line) and 10 MHz (dashed line).}
\label{fig:Fig4}
\end{figure*}

We further examine the interplay of spatial frequencies of the heater and the overall thermal response in Fig.~\ref{fig:Fig4}. In Fig.~\ref{fig:Fig4}(a), the effective thermal conductivity is plotted versus period for two duty cycles. The non-monotonic dependence of the effective thermal conductivity is again observed. While the period sets the relevant spatial frequency as $2n\pi/L$ where $n$ is integer, the duty cycle determines the amplitude of each term of the sum in the heating input function of a dot array. In both cases, a dip in effective thermal conductivity is observed as 
$\xi_r \Lambda_{\omega} \sim 1$. However, the amplitude of the DC component at 90 \% duty cycle relative to the first harmonic is 81 times higher than that at 10 \% duty cycle, which explains why the duty cycle at 10 \% results in a bigger drop in the effective thermal conductivity. As the period keeps decreasing, the DC component becomes more dominant in both cases, leading to the increase in the effective thermal conductivity. 

Figure~\ref{fig:Fig4}(b) demonstrates the interplay between the cross-plane and in-plane effects. As the temporal frequency increases, cross-plane effects become more apparent, and the effective thermal conductivity converges to a smaller value at large periods. Interestingly, the drop of the thermal conductivity becomes more prominent as the increase of the cross-plane effects, and the minimum shifts to a smaller period. The origin of this behavior is that the DC component scales with the temporal frequency as $Q\eta^{-1}C^{-1}$ while the higher harmonics have a much weaker dependence on $\eta$. Therefore, as the temporal frequency increases, the relative amplitude of the harmonics compared to the DC component also increases, and the drop in thermal conductivity becomes bigger. Also, as the temporal frequency increases, the relevant range of $\xi_z$ scales up as $\sqrt{k\eta^{-1}C^{-1}}$, accounting for the shift of the dip to smaller in-plane spatial frequencies in Fig.~\ref{fig:Fig4}(b) as temporal frequency increases.

\section{Conclusion}

In summary, we have studied quasiballistic transport from various heater geometries using an analytical solution of the Boltzmann equation in the spatial frequency domain. We find that the effective thermal resistance of nanoscale heat sources depends on several factors and is not necessarily different from the Fourier value. Whether a difference occurs depends on the relevant spatial frequencies of the heater, the amplitude of the DC component relative to those of other spatial frequencies, the spectral profile of the heat source, and the temporal frequency. Our work provides detailed insights into the thermal resistance in the quasiballistic regime, a finding that could impact strategies for thermal management in electronics and other applications.

\bibliographystyle{unsrt}

\begin{thebibliography}{10}

\bibitem{Chen}
G.~Chen.
\newblock {\em Nanoscale Energy Transport and Conversion}.
\newblock Oxford University Press, Inc, 2005.

\bibitem{Majumdar1993}
A.~Majumdar.
\newblock Microscale heat conduction in dielectric thin-films.
\newblock {\em J. heat transfer}, 115:7--16, 1993.

\bibitem{hua_transport_2014}
Chengyun Hua and Austin~J. Minnich.
\newblock Transport regimes in quasiballistic heat conduction.
\newblock {\em Phys. Rev. B}, 89:094302, Mar 2014.

\bibitem{Cahill:2014}
David~G. Cahill, Paul~V. Braun, Gang Chen, David~R. Clarke, Shanhui Fan,
  Kenneth~E. Goodson, Pawel Keblinski, William~P. King, Gerald~D. Mahan, Arun
  Majumdar, Humphrey~J. Maris, Simon~R. Phillpot, Eric Pop, and Li~Shi.
\newblock {Nanoscale thermal transport. {II.} 2003--2012}.
\newblock {\em Applied Physics Reviews}, 1(1):011305, January 2014.

\bibitem{Siemens2010}
Mark~E. Siemens, Qing Li, Ronggui Yang, Keith~A. Nelson, Erik~H. Anderson,
  Margaret~M. Murnane, and Henry~C. Kapteyn.
\newblock Quasi-ballistic thermal transport from nanoscale interfaces observed
  using ultrafast coherent soft x-ray beams.
\newblock {\em Nat. Mater.}, 9:26--30, 2010.

\bibitem{Johnson2013}
Jeremy~A. Johnson, A.~A. Maznev, John Cuffe, Jeffrey~K. Eliason, Austin~J.
  Minnich, Timothy Kehoe, Clivia M.~Sotomayor Torres, Gang Chen, and Keith~A.
  Nelson.
\newblock Direct measurement of room-temperature nondiffusive thermal transport
  over micron distances in a silicon membrane.
\newblock {\em Phys. Rev. Lett.}, 110:025901, Jan 2013.

\bibitem{Minnich2011PRL}
A.~J. Minnich, J.~A. Johnson, A.~J. Schmidt, K.~Esfarjani, M.~S. Dresselhaus,
  K.~A. Nelson, and G.~Chen.
\newblock Thermal conductivity spectroscopy technique to measure phonon mean
  free paths.
\newblock {\em Phys. Rev. Lett.}, 107:095901, Aug 2011.

\bibitem{ding_radial_2014}
D.~Ding, X.~Chen, and A.~J. Minnich.
\newblock Radial quasiballistic transport in time-domain thermoreflectance
  studied using monte carlo simulations.
\newblock {\em Applied Physics Letters}, 104(14):143104.

\bibitem{Cahill2007}
Yee~Kan Koh and David~G. Cahill.
\newblock Frequency dependence of the thermal conductivity of semiconductor
  alloys.
\newblock {\em Phys. Rev. B}, 76:075207, Aug 2007.

\bibitem{Wilson2014}
R.~B. Wilson and David~G. Cahill.
\newblock Anisotropic failure of fourier theory in time-domain
  thermoreflectance experiments.
\newblock {\em Nature Communications}, 5:5075, 2014.

\bibitem{zeng_measuring_2015}
Lingping Zeng, Kimberlee~C. Collins, Yongjie Hu, Maria~N. Luckyanova, Alexei~A.
  Maznev, Samuel Huberman, Vazrik Chiloyan, Jiawei Zhou, Xiaopeng Huang,
  Keith~A. Nelson, and Gang Chen.
\newblock Measuring phonon mean free path distributions by probing
  quasiballistic phonon transport in grating nanostructures.
\newblock {\em Scientific Reports}, 5:17131.

\bibitem{hu_spectral_2015}
Yongjie Hu, Lingping Zeng, Austin~J. Minnich, Mildred~S. Dresselhaus, and Gang
  Chen.
\newblock Spectral mapping of thermal conductivity through nanoscale ballistic
  transport.
\newblock {\em Nature Nanotechnology}, 10(8):701--706.

\bibitem{hoogeboom-pot_new_2015}
Kathleen~M. Hoogeboom-Pot, Jorge~N. Hernandez-Charpak, Xiaokun Gu, Travis~D.
  Frazer, Erik~H. Anderson, Weilun Chao, Roger~W. Falcone, Ronggui Yang,
  Margaret~M. Murnane, Henry~C. Kapteyn, and Damiano Nardi.
\newblock A new regime of nanoscale thermal transport: Collective diffusion
  increases dissipation efficiency.
\newblock {\em Proceedings of the National Academy of Sciences},
  112(16):4846--4851.

\bibitem{Chen2017}
Xiangwen Chen, Chengyun Hua, Navaneetha~K. Ravichandran, and Austin~J. Minnich.
\newblock Thermal response of materials to extreme temperature gradients and
  the role of the spatial frequency.
\newblock {\em under review}.

\bibitem{Vermeersch2015a}
Bjorn Vermeersch, Amr M.~S. Mohammed, Gilles Pernot, Yee~Rui Koh, and Ali
  Shakouri.
\newblock Superdiffusive heat conduction in semiconductor alloys. ii. truncated
  l\'evy formalism for experimental analysis.
\newblock {\em Phys. Rev. B}, 91:085203, Feb 2015.

\bibitem{Vermeersch2015b}
Bjorn Vermeersch, Jes\'us Carrete, Natalio Mingo, and Ali Shakouri.
\newblock Superdiffusive heat conduction in semiconductor alloys. i.
  theoretical foundations.
\newblock {\em Phys. Rev. B}, 91:085202, Feb 2015.

\bibitem{Maznev2011}
A.~A. Maznev, Jeremy~A. Johnson, and Keith~A. Nelson.
\newblock Onset of nondiffusive phonon transport in transient thermal grating
  decay.
\newblock {\em Phys. Rev. B}, 84:195206, Nov 2011.

\bibitem{Wilson2013}
R.~B. Wilson, Joseph~P. Feser, Gregory~T. Hohensee, and David~G. Cahill.
\newblock Two-channel model for nonequilibrium thermal transport in pump-probe
  experiments.
\newblock {\em Phys. Rev. B}, 88:144305, Oct 2013.

\bibitem{Dames2015}
Fan Yang and Chris Dames.
\newblock Heating-frequency-dependent thermal conductivity: An analytical
  solution from diffusive to ballistic regime and its relevance to phonon
  scattering measurements.
\newblock {\em Phys. Rev. B}, 91:165311, Apr 2015.

\bibitem{Maassen2015}
Jesse Maassen and Mark Lundstrom.
\newblock Steady-state heat transport: Ballistic-to-diffusive with fourier's
  law.
\newblock {\em Journal of Applied Physics}, 117(3):035104, 2015.

\bibitem{Regner2014}
K.~T. Regner, A.~J.~H. McGaughey, and J.~A. Malen.
\newblock Analytical interpretation of nondiffusive phonon transport in
  thermoreflectance thermal conductivity measurements.
\newblock {\em Phys. Rev. B}, 90:064302, Aug 2014.

\bibitem{Ramu2014}
Ashok~T. Ramu and Yanbao Ma.
\newblock An enhanced fourier law derivable from the boltzmann transport
  equation and a sample application in determining the mean-free path of
  nondiffusive phonon modes.
\newblock {\em Journal of Applied Physics}, 116:093501, 2014.

\bibitem{Collins2013APL}
Kimberlee~C. Collins, Alexei~A. Maznev, Zhiting Tian, Keivan Esfarjani,
  Keith~A. Nelson, and Gang Chen.
\newblock Non-diffusive relaxation of a transient thermal grating analyzed with
  the boltzmann transport equation.
\newblock {\em Journal of Applied Physics}, 114(10):--, 2013.

\bibitem{zeng_disparate_2014}
Lingping Zeng and Gang Chen.
\newblock Disparate quasiballistic heat conduction regimes from periodic heat
  sources on a substrate.
\newblock {\em Journal of Applied Physics}, 116(6):064307.

\bibitem{Lindsay2013}
L.~Lindsay, D.~A. Broido, and T.~L. Reinecke.
\newblock Ab initio thermal transport in compound semiconductors.
\newblock {\em Phys. Rev. B}, 87:165201, Apr 2013.

\bibitem{hua_analytical_2014}
Chengyun Hua and Austin~J. Minnich.
\newblock Analytical green's function of the multidimensional
  frequency-dependent phonon boltzmann equation.
\newblock {\em Phys. Rev. B}, 90:214306, Dec 2014.

\bibitem{Hua2017}
Chengyun Hua, Xiangwen Chen, Navaneetha~K. Ravichandran, and Austin~J. Minnich.
\newblock Experimental metrology to obtain thermal phonon transmission
  coefficients at solid interfaces.
\newblock {\em Phys. Rev. B}, 95:205423, May 2017.

\end{thebibliography}

\end{document}